\documentclass[showpacs,fleqn,nobibnotes]{revtex4}

\usepackage{amsmath}
\usepackage{graphicx}
\usepackage{float}

\def\lsim{\raise0.3ex\hbox{$<$\kern-0.75em\raise-1.1ex\hbox{$\sim$}}}
\def\gsim{\raise0.3ex\hbox{$>$\kern-0.75em\raise-1.1ex\hbox{$\sim$}}}

\newcommand{\rr}{\mbox{\boldmath $r$}}

\newcommand{\rb}{\mbox{\boldmath $b$}}

\newcommand{\be}{\begin{equation}}
\newcommand{\ee}{\end{equation}}

\begin{document}

\title{Heavy quark production and gluon saturation in double parton scattering at LHC}
\author{E.R.  Cazaroto$^1$ , V.P. Gon\c{c}alves$^2$ and F.S. Navarra$^1$}

\affiliation{$^1$ Instituto de F\'{\i}sica, Universidade de S\~{a}o Paulo,
C.P. 66318,  05315-970 S\~{a}o Paulo, SP, Brazil\\
 $^2$ Instituto de F\'{\i}sica e Matem\'atica, Universidade Federal de
Pelotas\\
Caixa Postal 354, CEP 96010-900, Pelotas, RS, Brazil.}

\begin{abstract}
The production of $c\bar{c}c\bar{c}$, $b\bar{b}b\bar{b}$ and $c\bar{c}b\bar{b}$ pairs considering   double parton scattering   at LHC energies is investigated. 
We estimate the contribution of saturation effects to the different final states and predict the energy dependence of the cross sections.  
Moreover, we estimate the ratio between the double and single parton scattering cross sections for the full rapidity range of the LHC and for the rapidity range 
of the LHCb experiment. For the full rapidity range we confirm a previous prediction, namely that for charm production the double parton scattering 
contribution  becomes comparable with the single parton scattering one at  LHC energies. We also  demonstrate that this result remains valid when one introduces 
saturation effects in the calculations. 
Finally  we show that the production of $c\bar{c}b\bar{b}$ contributes significantly to  bottom production.  For the LHCb kinematical range the ratio is strongly 
reduced. 

\end{abstract}

\maketitle

\section{Introduction}

Heavy quark production in hard collisions of hadrons, leptons, and photons has been considered as a clean test of
perturbative QCD. This process provides not only many tests of perturbative QCD, but also some of the most important backgrounds to new physics processes, 
which have motivated a comprehensive phenomenological studies carried out  at DESY-HERA, Tevatron and LHC. The study of heavy quark production also 
is  motivated by the strong dependence of
the cross section on the behaviour of the gluon distribution, which determines the QCD dynamics at high energies.
The huge density of low-$x$ gluons in the hadron wave-functions is expected to modify the usual description of the gluon distribution in terms of the linear 
DGLAP dynamics \cite{dglap} by the inclusion of non-linear corrections associated to the physical process of parton recombination. In particular, it is 
expected the 
formation of a Color Glass Condensate (CGC) \cite{cgc}, which  is characterized by the limitation on the maximum phase-space parton density that can be 
reached in the hadron
wave-function (parton saturation) and described in the mean-field approximation by the Balitsky - Kovchegov (BK) equation \cite{bk}.  These saturation effects  
are expected to contribute significantly at high energies leading to the breakdown of the twist expansion and of the factorization schemes (For recent reviews 
see Ref. \cite{hdqcd}). In Ref. \cite{hqp_nos} we studied the impact of the saturation effects in the single  heavy quark pair  production in proton-proton and 
proton-nucleus collisions at LHC energies considering  the color dipole formalism and the solution of the running coupling 
Balitsky-Kovchegov equation, which is currently the state-of-art of the CGC formalism. One of the goals of this paper is to compare our predictions with recent 
experimental data.

\begin{figure}[t]
\begin{tabular}{cc}
\includegraphics[scale=0.3]{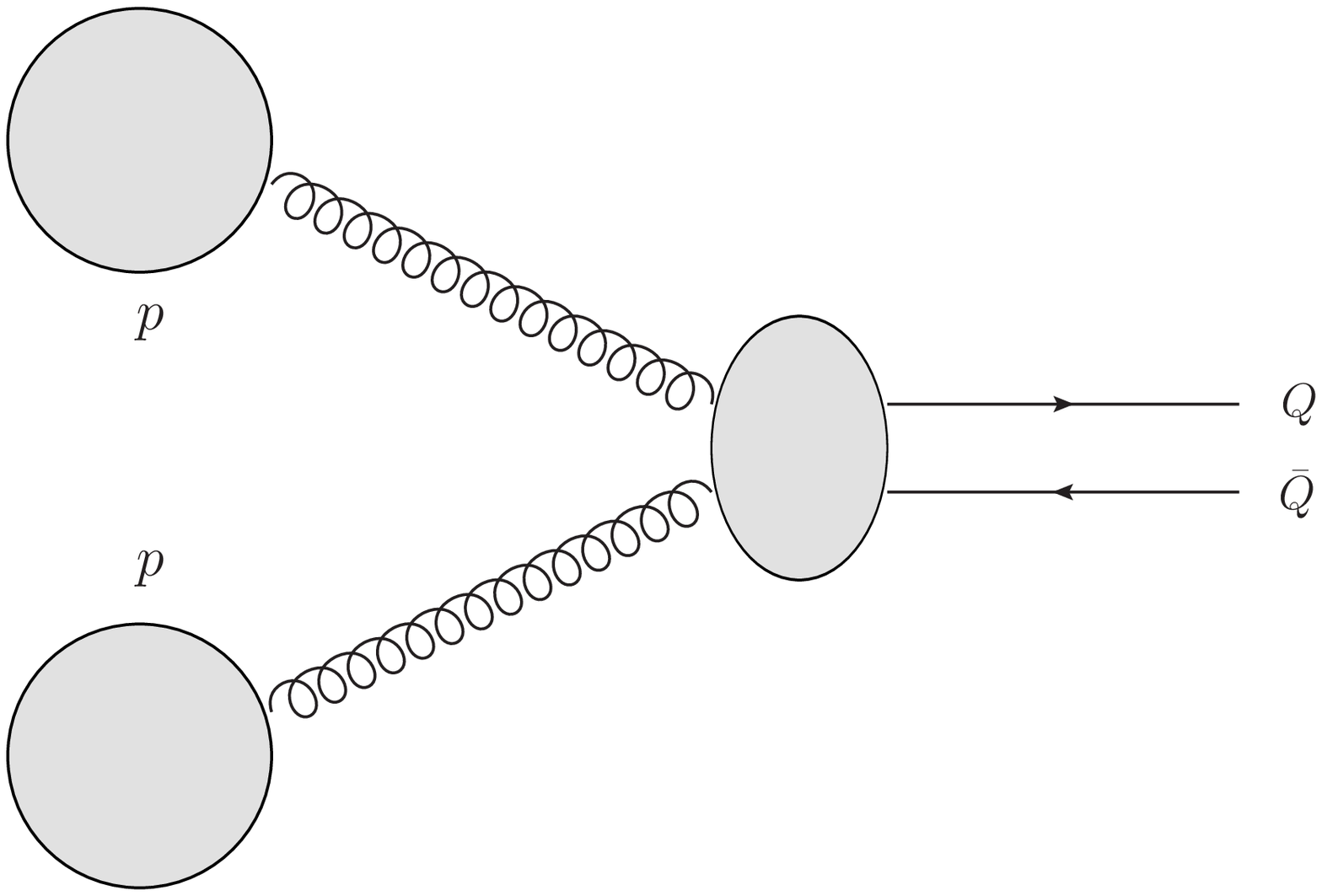}        
\includegraphics[scale=0.3]{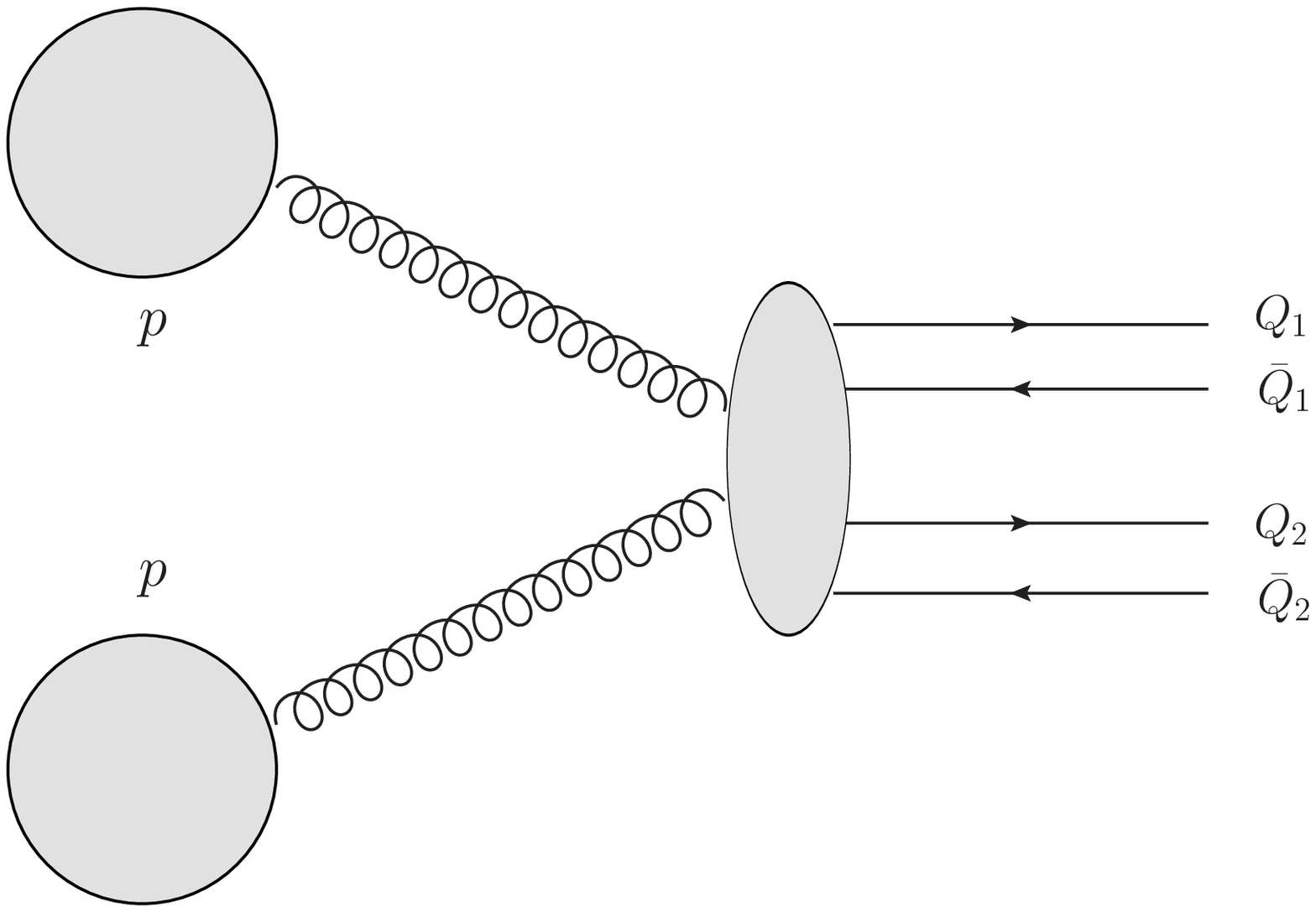}        
\includegraphics[scale=0.3]{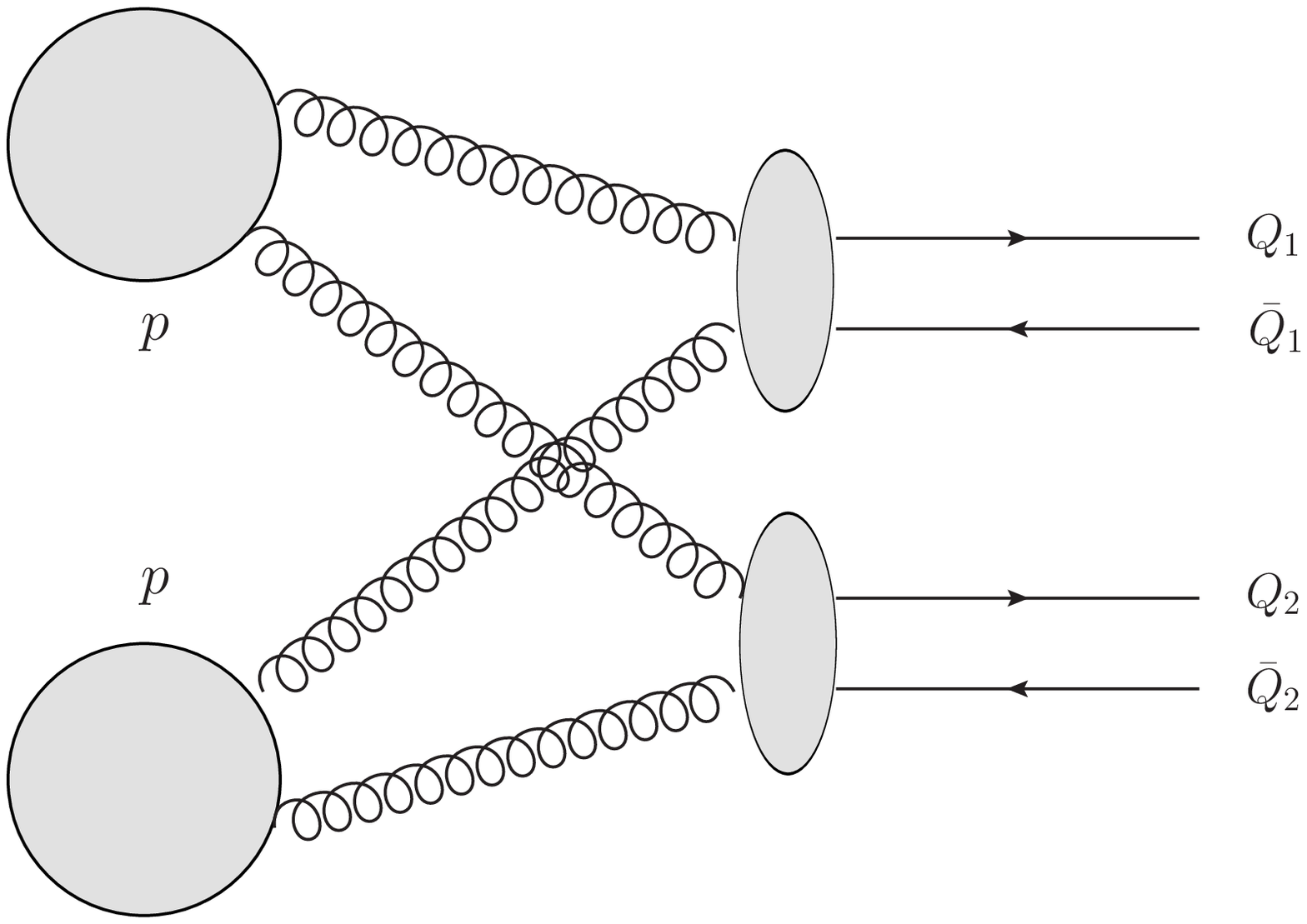}        
\end{tabular}
\caption{(Color online) Left: The $Q\bar{Q}$ pair production in the single parton scattering (SPS) process; Center: The  $Q_1\bar{Q}_1Q_2\bar{Q}_2$ pair
 production in the SPS process;  Right:   $Q_1\bar{Q}_1Q_2\bar{Q}_2$ pair production in the  double parton scattering (DPS) process.}
\label{fig:1}
\end{figure}

The high density of gluons in the initial state of hadronic collisions at LHC also implies that the probability of  multiple gluon - gluon interactions  within 
one proton - proton collision increases. In particular, the probability of having two or more hard interactions in a collision is not significantly 
suppressed with respect to the single interaction probability. It has motivated a rapid development of the theory of double parton scattering (DPS) processes 
 and several estimates of the cross sections for different processes have been presented in recent years. In particular, the production of two $c\bar{c}$ pairs 
in double-parton scattering was discussed recently  in Ref. \cite{Marta_Rafal} (See also Ref. \cite{liko_prd86}), which obtained the 
surprising result that at the energies of LHC the contribution of the DPS channel for two $c\bar{c}$ pairs production  [See Fig. \ref{fig:1} (right)]  becomes 
of the same order of the single parton scattering (SPS) channel contribution for one pair production [See Fig. \ref{fig:1} (left)], with the production of  two
 $c\bar{c}$ pairs   in SPS processes [See Fig. \ref{fig:1} (center)] being  strongly suppressed.  Another of the goals of this paper is to complement the study 
performed in Ref. \cite{Marta_Rafal} by the inclusion of saturation effects and by the analysis of the two  $b\bar{b}$ pairs production. Moreover, we estimate 
for the first time the cross section for  the  $c\bar{c} b\bar{b}$ production in DPS processes.

This paper is organized as follows. In the next Section (Section \ref{heavy}) we present a brief review of heavy quark production in SPS and DPS processes. 
In Section
\ref{dipole} we present the main formulae for the calculation of the  one pair $Q\bar{Q}$ production in the color dipole formalism. We also make a brief review 
of how to include saturation effects in the color dipole formalism and present the models that we will use in the calculations.  In section \ref{resultados} we  
present our predictions for the energy dependence of the   $c\bar{c} c\bar{c}$, $b\bar{b} b\bar{b}$ and $c\bar{c} b\bar{b}$ production cross sections. Finally, 
in Section \ref{conc} we summarize our main results and conclusions.

\section{Heavy quark production}
\label{heavy}

The calculation of the heavy quark cross section in the standard framework assumes that only one hard interaction occurs per collision. 
This mechanism is called single-parton scattering (SPS), since the Feynman diagram contains one gluon from the hadron target and one gluon from the hadron 
projectile [See Fig. \ref{fig:1} (left)]. The next-to-leading order (NLO) correction for this process was already studied \cite{nason0,alta,nason1,been}. 
In general, higher order corrections do not change significantly the observables, since the contributions are suppressed by powers of $\alpha_s$. For example,
 the  $Q_1\bar{Q}_1Q_2\bar{Q}_2$ ($Q_i = c$ or $b$)  production in SPS processes [See Fig. \ref{fig:1} (center)] is suppressed by a factor proportional to
 $ \alpha_s^4$. The basic idea, which justifies the SPS approach, is that the probability of a hard  interaction in a collision is very small, which makes the 
probability of having two or more hard interactions in a collision highly suppressed with respect to the single interaction probability. Such assumption is 
reasonable in the kinematical regime in which the flux of incoming partons is not very high. However, as already pointed in Ref. \cite{liko_prd86}, at LHC 
energies the hadronic cross section appears to be three orders of magnitude higher than the cross section of the partonic subprocess. In this condition, 
there is a high probability of scattering of more than one pair of partons in the same hadron - hadron collision. This expectation has been recently confirmed  
by the LHCb Collaboration \cite{lhcb_jhep}, which has observed the production of $J/\Psi$ mesons accompanied by open charm and pairs of open charm 
hadrons in $pp$ collisions at $\sqrt{s} = 7$ TeV and verified that the SPS predictions are significantly smaller than the observed cross sections. Furthermore, 
in Ref. \cite{rafal} the authors demonstrated that the description of the inclusive ATLAS, ALICE and LHCb data is very difficult only in terms of the SPS 
contribution.

Following the same factorization approximations assumed for processes with a single hard scattering, it is possible to derive the DPS contribution for the heavy 
quark cross section considering two independent hard parton sub-processes. It is given  by  (See, e.g. Ref. \cite{diehl_jhep})
\begin{eqnarray}
\sigma_{h_1 h_2 \rightarrow Q_1\bar{Q}_1Q_2\bar{Q}_2}^{DPS} = \left( \frac{m}{2} \right)  \int \Gamma_{h_1}^{gg} (x_1,x_2; \rb_1, \rb_2; \mu_1^2, \mu_2^2) 
\hat{\sigma}_{Q_1\bar{Q}_1}^{gg} (x_1,x_1^{\prime},\mu_1^2) \hat{\sigma}_{Q_2\bar{Q}_2}^{gg} (x_2,x_2^{\prime},\mu_2^2) \nonumber \\
\times  \Gamma_{h_2}^{gg} (x_1^{\prime},x_2^{\prime};\rb_1-\rb,\rb_2-\rb;\mu_1^2,\mu_2^2) dx_1 dx_2 dx_1^{\prime} dx_2^{\prime} d^2b_1 d^2b_ 2 d^2b \,\,,
\label{sigdps_geral}
\end{eqnarray}
where we assume that the quark-induced sub-processes can be disregarded at high energies,   $\Gamma_{h_1}^{gg} (x_1,x_2; \rb_1, \rb_2; \mu_1^2, \mu_2^2)$ are 
the two gluon parton distribution functions which depend on the longitudinal momentum fractions $x_1$ and $x_2$, and on the transverse position $\rb_1$ and 
$\rb_2$ of the two gluons undergoing the hard processes at the scales $\mu_1^2$ and $\mu_2^2$. The functions $\hat{\sigma}$ are the parton level sub-processes
 cross sections and $\rb$ is the impact parameter vector connecting the centers of the colliding protons in the transverse plane. Moreover, $m/2$ is a 
combinatorial factor which accounts  for  indistinguishable  and distinguishable  final states. For $Q_1 = Q_2$ one has $m=1$, while $m=2$ for $Q_1 \neq Q_2$.  
It is common in the literature to assume that the longitudinal and transverse components of the  double parton distributions can be decomposed   and that the 
longitudinal components can be expressed in terms of the product of two independent single parton distributions. The  proof of these assumptions in the general 
case is still  an open question (See, e.g. Refs. \cite{diehl_jhep,diehl_plb}).  In the particular case of heavy quark production, in Ref. \cite{Marta_Rafal} the
 authors compared the results of this simple factorized Ansatz with those obtained using double parton distributions with QCD evolution and  verified that the 
predictions are similar if we taken  into account all uncertainties present in the calculations as, for instance, those associated to the choice of the
 factorization and renormalization scales. In the present study we  will also assume the validity of these assumptions and consider that the DPS contribution to 
 the heavy quark cross section can be expressed in a simple generic form given by
\begin{eqnarray}
\sigma_{h_1 h_2 \rightarrow Q_1\bar{Q}_1Q_2\bar{Q}_2}^{DPS} = \left( \frac{m}{2} \right) \frac{\sigma^{SPS}_{h_1 h_2 \rightarrow Q_1\bar{Q}_1} 
\sigma^{SPS}_{h_1 h_2 \rightarrow Q_2\bar{Q}_2}}{\sigma_{eff}} \,\,,
\label{dps_fac}
\end{eqnarray}
where $\sigma_{eff}$ is a normalization cross section representing the effective transverse overlap of partonic interactions that produce the DPS process.  
As in \cite{Marta_Rafal} we assume $\sigma_{eff} = 15$ mb.
This formula expresses  the DPS cross section as the product of two individual SPS cross sections assuming that the two SPS sub-processes are uncorrelated and do 
not interfere.

A comment is in order. In what follows we will estimate the DPS cross section using Eq. (\ref{dps_fac}) and taking into account saturation effects in the SPS 
cross section, which will be discussed in the next section. We are aware that Eq. (\ref{dps_fac}) may not be valid when the saturation effects become important. 
However, we believe that in the particular case of heavy quark production at LHC energies it allows us to obtain a reasonable first approximation for the magnitude 
of these effects in the DPS process.

\section{Saturation effects in  heavy quark production}
\label{dipole}

Saturation effects can be naturally described in the color dipole formalism. At high energies color dipoles with a defined 
transverse  separation are  eigenstates of the interaction. The main quantity in this formalism is the dipole-target cross section, which is 
universal and determined by  QCD dynamics at high energies. In particular, it provides an unified description of inclusive and diffractive observables 
in $ep$ processes as well as of   Drell-Yan pairs, prompt photon and heavy quark production in hadron-hadron collisions. 

The description of  heavy quark production in the color dipole formalism was  proposed in Refs. \cite{Nikzak1,Nikzak2} and discussed in detail in Refs. 
\cite{kope_tarasov,rauf} (See also Refs. \cite{Tuchin1,Tuchin2,Tuchin3}). The basic idea is the following. Before interacting with the hadron target $h_2$ a 
gluon is emitted by the hadron projectile $h_1$, which fluctuates into  a color octet pair $Q\bar{Q}$. In the low-$x$ regime the time of fluctuation is much larger
 than the time of interaction, and color dipoles with a defined transverse separation $\vec{\rho}$ are eigenstates of the interaction. Consequently, the total 
cross section for the process $h_1 h_2 \rightarrow Q\bar{Q} X$ is then given by \cite{Nikzak1,Nikzak2}:
\begin{equation}
\sigma (h_1 h_2 \rightarrow \{ Q\bar{Q} \}X) = 
2 \int _0 ^{-ln(2m_Q/ \sqrt{s})} dy  \, x_1 \, G_{h_1}(x_1,\mu _F) \,
\sigma (Gh_2 \rightarrow \{Q\bar{Q}\} X)
\label{sigtot}
\end{equation}
where $x_1G_{h_1}(x_1,\mu _F)$ is the projectile gluon distribution, the cross section  $\sigma (Gh_2 \rightarrow \{Q\bar{Q}\} X)$ describes 
the heavy quark production in the gluon - target interaction, $y$  is the rapidity of the pair and $\mu_F$ is the 
factorization  scale. The cross section for the process $G + h_2 \rightarrow Q \bar{Q} X$ is given by:
\begin{equation}
\sigma(G h_2 \rightarrow\{Q\bar{Q}\}X) = \int _0^1 d \alpha \int d^2\rho \,\, 
\vert \Psi _{G\rightarrow Q\bar{Q}} (\alpha,\rho)\vert ^2 
\,\, \sigma^{h_2} _{Q\bar{Q}G}(\alpha , \rho)
\label{sec1}
\end{equation}
where $  \sigma^{h_2}_{Q\bar{Q}G}$  is the scattering cross section of a color neutral quark-antiquark-gluon system on the 
hadron target $h_2$ \cite{Nikzak1,Nikzak2,kope_tarasov,rauf}: 
\begin{equation}
\sigma^{h_2}_{Q\bar{Q}G}(\alpha , \rho) = \frac{9}{8}[\sigma _{Q\bar{Q}}(\alpha \rho) + \sigma _{Q\bar{Q}}(\bar{\alpha} \rho)]
- \frac{1}{8}\sigma _{Q\bar{Q}}(\rho)\,\,. 
\label{sec2}
\end{equation}
The quantity $\sigma _{Q\bar{Q}}$ is the scattering cross section of a  color neutral quark-antiquark
pair with separation radius $\rho$ on the hadron target  and $\alpha$ ($\bar{\alpha} = 1 - \alpha$) is the 
fractional momentum of quark (antiquark). The light-cone (LC) wave-function of the 
transition $G \rightarrow  Q \bar{Q} $ can be calculated perturbatively, with the squared wave-function given by:
\begin{equation}
\vert \Psi _{G\rightarrow Q\bar{Q}} (\alpha,\rho)\vert ^2   
= \frac{\alpha _s (\mu _R)}{(2\pi)^2} \lbrace m^2_Q K_0^2(m_Q\rho)
+ [\alpha ^2 + \bar{\alpha ^2}] m^2_Q K_1^2(m_Q\rho)\rbrace
\label{psi}
\end{equation}
where $\alpha_s(\mu_R)$ is the strong coupling constant. Following Ref. \cite{hqp_nos} we will  assume that  $\mu_F=2 m_Q$ and that $xG$ is given in terms of the 
 GRV98 parton distribution \cite{grv}, but similar predictions are obtained using, e.g., the CTEQ6L parametrization \cite{cteq}.

In order to estimate the heavy quark cross section we need to specify the dipole - target cross section. 
In the Color Glass Condensate formalism  \cite{cgc} it  is given in terms of the dipole-target forward scattering amplitude ${\cal{N}}(x,\rho,\rb)$, which 
encodes all the
information about the hadronic scattering, and thus about the non-linear and quantum effects in the hadron wave function. It reads: 
\begin{eqnarray}
\sigma_{Q \bar{Q}} (x,\rho)=2 \int d^2 \rb \, {\cal{N}}(x,\rho,\rb)\,\,.
\end{eqnarray}
It is useful to assume that the impact parameter dependence of $\cal{N}$ can be factorized as 
${\cal{N}}(x,\rho,\rb) = {\cal{N}}(x,\rho) S(\rb)$, so that 
$\sigma_{Q \bar Q}(x,\rho) = {\sigma_0} \,{\cal{N}}(x,\rho)$, 
with $\sigma_0$ being a free parameter
related to the non-perturbative QCD physics. The Balitsky-JIMWLK hierarchy \cite{cgc,bk}  describes the energy evolution of the dipole-target
scattering amplitude ${\cal{N}}(x,\rho)$. In the mean field approximation, the first equation of this  hierarchy decouples and becomes  
the Balitsky-Kovchegov (BK) equation \cite{bk}.
However, an exact analytical solution to BK equation is unknown. A numerical solution, encoded in a FORTRAN subroutine, which considers running coupling 
corrections to the kernel of BK equation, is available in the literature \cite{bkrunning}. The calculations using this numerical solution will be denoted 
as `` rcBK '' hereafter. Currently, the rcBK model is the most sophisticated saturation model available in the literature.  
 We also present the predictions obtained using  the phenomenological saturation model proposed by Golec - Biernat and Wusthoff in Ref. \cite{gbw} (denoted GBW 
hereafter), in which the dipole - proton cross section is given by:
\begin{eqnarray}
\sigma^{GBW} _{Q \bar{Q}}  (x,\rho) = \sigma_0 \left[1-\exp(-\frac{\rho^2Q_{s}^2(x)}{4})\right],
\end{eqnarray}
where the saturation scale is given by $Q_{s}^2(x)=Q_0^2\left(x_0/x\right)^{\lambda}$, with $Q_0^2 = 1$ GeV$^2$, $x_0 =3\,.\,10^{-4} $ and $\lambda = 0.288$. 
Our motivation to use this model, which has been updated in several aspects in the last years, is that it allows us to easily  obtain its linear limit,  given by  
$\sigma^{GBW\,linear} _{Q \bar{Q}} = \sigma_0 {\rho^2Q_{s}^2(x)}/{4}$. Consequently, it allows to quantify the contribution 
of the saturation effects in the observable under analysis.

\section{Results and Discussion}
\label{resultados}

\begin{figure}[t]
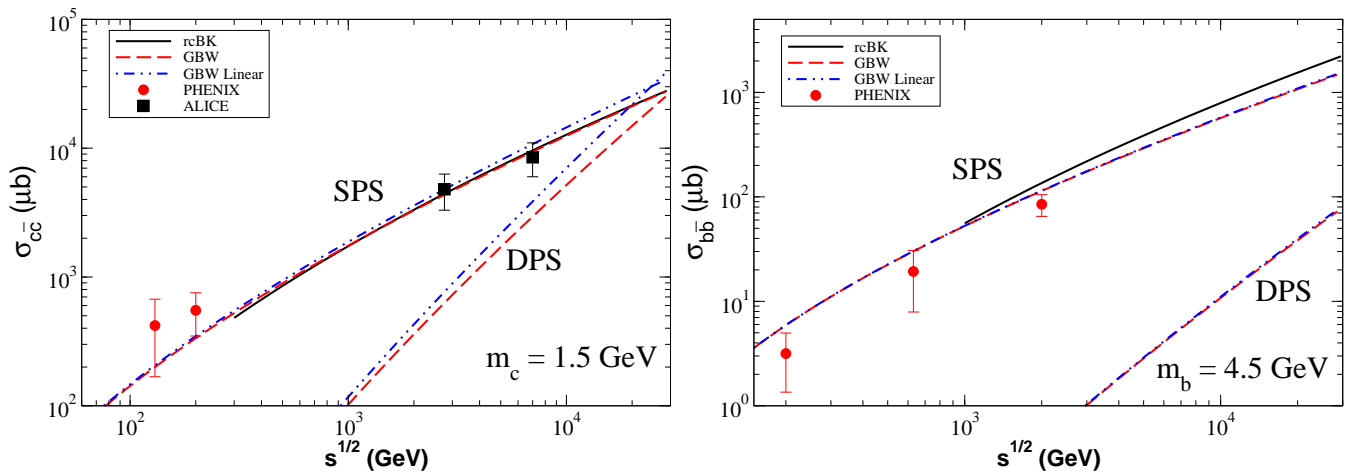

\begin{tabular}{cc}
\includegraphics[scale=0.35]{sig_tot_e_dps_charm.eps}        
\includegraphics[scale=0.35]{sig_tot_e_dps_bottom.eps}        
\end{tabular}
\caption{(Color online) Charm (left) and bottom (right) production cross sections in Single Parton Scattering (SPS) and Double Parton Scattering (DPS) as a 
function of the c.m.s. energy ($\sqrt{s}$). Data points from PHENIX \cite{adcox02,adare06} (circles) and  from  ALICE \cite{ALICE_pp} (squares).}
\label{fig:2}
\end{figure}

The heavy quark production in SPS processes considering saturation effects was studied in detail in Ref. \cite{hqp_nos}. There we predicted the energy dependence 
of the  charm and bottom pair production and compared with data points from UA2, PHENIX and from Cosmic Rays. All these data can be quite well described using the 
color dipole formalism and an adequate choice of the  heavy quark mass. Recently, the ALICE Collaboration has released its first data for charm production at 
$\sqrt{s}=2.76$ TeV  \cite{ALICE_pp}. It allows us to compare, for the first time, the color dipole formalism for heavy quark production in hadronic collisions 
with experimental data at high energies, which is the kinematical range where it is theoretically justified.  In Fig. \ref{fig:2} (left) we compare the rcBK, 
GBW and GBW Linear predictions with the ALICE \cite{ALICE_pp} and PHENIX \cite{adcox02,adare06} data considering $m_c = 1.5$ GeV.  We can see that the different 
models for the dipole-target cross sections are able to describe the experimental data. 
The rcBK and GBW predictions are almost identical  in the whole range of energy. When  the GBW Linear model is used as input in the calculations, we predict larger 
values for the charm cross section. In contrast, for bottom production [See Fig. \ref{fig:2} (right)],  the GBW and GBW Linear predictions are identical in the 
considered energy range and the rcBK one predicts larger values of the cross section. Unfortunately, up to now, LHC data for bottom production are not available.
The distinct behavior observed for charm and bottom production is directly associated to the fact that in the color dipole formalism the contribution of the 
non-linear effects is determined by the integrand of the pair separation ($\rho$)  integral [See Eq. (\ref{sec1})], i.e. by the product of the wave-function 
squared and the dipole-target cross section. This integrand has a peak at $\rho \approx 1/m_Q$. Consequently, for bottom production, the integral is dominated 
by very small pair separations, probing the linear regime of the dipole-target cross section. The rcBK prediction is larger than the GBW and GBW Linear ones
 because its linear regime is associated to the BFKL dynamics, which implies a steep energy dependence. In contrast, for charm production, we probe larger 
values of the pair separation, where saturation effects cannot be disregarded. The difference between the rcBK and the GBW predictions is associated to the 
delayed saturation predicted by the rcBK equation.

\begin{figure}[t]
\begin{tabular}{cc}
\includegraphics[scale=0.35]{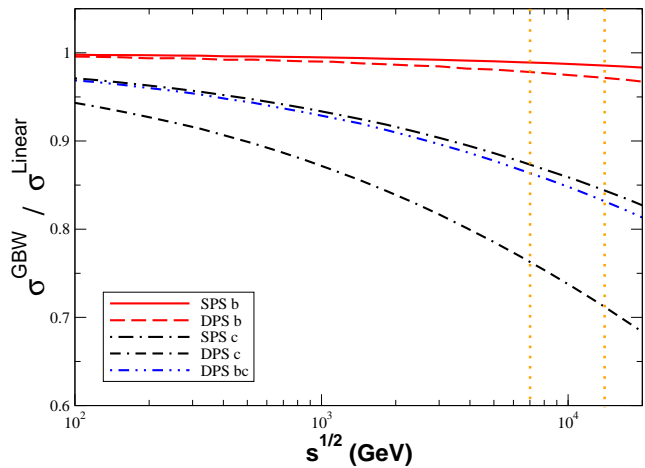}
\end{tabular}
\caption{(Color online) The ratio $\, \sigma^{GBW} / \sigma^{GBW Linear} \,$ for charm and bottom production in SPS and DPS processes as a function of the c.m.s.
 energy ($\sqrt{s}$).}
\label{fig:3}
\end{figure}

In Fig. \ref{fig:2} we also present the $c\bar{c}c\bar{c}$ and $b\bar{b}b\bar{b}$ production cross sections in DPS processes considering only the GBW and GBW Linear 
models, for simplicity. 
For $c\bar{c}c\bar{c}$ production, we confirm the conclusion from \cite{Marta_Rafal},  that DPS charm production cross section becomes comparable with the SPS one 
at  LHC energies. 
This result remains valid when one considers saturation effects in the calculations. On the other hand,  the $b\bar{b}b\bar{b}$ production in DPS processes is always
 negligible when compared to $b\bar{b}$ production in SPS processes.
In order to estimate the contribution of the saturation effects in these processes we calculate the ratio between the GBW and GBW Linear cross sections. In Fig. 
\ref{fig:3} we present the energy dependence of this ratio, which is equal approximately one when the saturation effects are small. The two vertical lines delimit
 the energy range $7 \le \sqrt{s} \le 14$ TeV.
For $b\bar{b}$ production (denoted SPS b in the figure) we can see that the magnitude of saturation effects is really very small in the energies of LHC. On the 
other hand, for $c\bar{c}$ production,  the saturation effects decrease the SPS cross section in $\approx 15 \%$. In the case of DPS processes, these effects are 
very small in the bottom case but are $\approx 28 \%$ in the $c\bar{c}c\bar{c}$ production.
 In Fig. \ref{fig:3} we also present the magnitude of the saturation effects for a third type of event: the $c\bar{c}b\bar{b}$ production in DPS processes. In this 
process, instead of two pairs of the same flavor we have the production of one $c\bar{c}$ pair and one $b\bar{b}$ pair. As we can see, the saturation effects in 
this type of process (denoted ``DPS bc'' hereafter) causes almost the same decrease ($\approx 15 \%$) that it causes in the SPS $c\bar{c}$ production. This almost 
identical decrease is a consequence of the fact that the $c\bar{c}b\bar{b}$ production cross section in DPS processes is given by the product of two SPS cross 
sections, one for $c\bar{c}$ production and one for $b\bar{b}$ production. Since $c\bar{c}$ production is much more sensitive to saturation effects than $b\bar{b}$ 
production, the saturation effects in $c\bar{c}b\bar{b}$ production come predominantly from the $c\bar{c}$ sector. Having discussed the magnitude of the saturation 
effects, in the following analysis we will only use the GBW model as input in our calculations.

\begin{figure}[t]
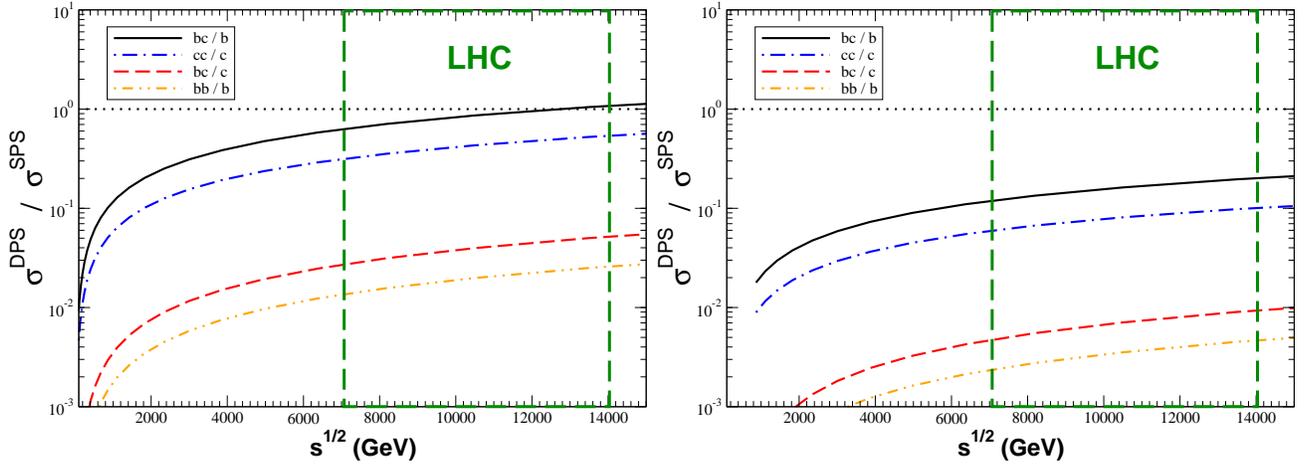

\begin{tabular}{cc}
\includegraphics[scale=0.35]{razoes_dps_por_sps.eps}
\includegraphics[scale=0.35]{razoes_dps_por_sps_lhcb.eps}
\end{tabular}
\caption{(Color online) The ratio $\sigma ^{DPS}/\sigma ^{SPS}$ as a function of the c.m.s. energy ($\sqrt{s}$) considering the full rapidity range (left) and the 
LHCb rapidity range (right).}
\label{fig:4}
\end{figure}

In Fig. \ref{fig:4} we present our predictions for the energy dependence of the ratio $\sigma ^{DPS}/\sigma ^{SPS}$. We denote by `` bc/b ''  the ratio between the
 cross sections for the  $b\bar{b}c\bar{c}$ production in DPS processes and for the $b\bar{b}$ production in SPS processes, and so on. 
As in previous figure, the vertical lines delimit the energy range probed by the LHC. In the left panel we present the results obtained integrating the cross
 sections in the full LHC rapidity range, while in the right panel the cross sections have been integrated in the LHCb rapidity range  $2 < y < 4.5$.
Considering initially the full LHC rapidity range, we have that for $\sqrt{s} = 7$ TeV the DPS charm production cross section is already of the same order of
 magnitude of the SPS charm production cross section, with the first reaching $\approx 30 \%$ of the value of the second. For $\sqrt{s} = 14$ TeV, this value
 reaches $\approx 60\%$.
In contrast, the ratio `` bb/b '' is almost 2$\%$ in the LHC energy range.
A surprising result is observed when we consider the ratio  
`` bc/b ''. We obtain that this ratio is $\, \approx 0.6 \,$ for $\sqrt{s} = 7$ TeV and $\, \approx 1 \,$ for $\sqrt{s} = 14$ TeV. It means that in $pp$
 collisions at $\sqrt{s} = 14$ TeV half the total amount of $b\bar{b}$ pairs produced in LHC will come from the DPS channel. 
When we consider the ratio for the restricted rapidity range probed by LHCb, we obtain that all predictions are significantly reduced, being always smaller than 
20 $\%$. In particular, the ratio `` bc/b '' in LHCb assumes the value $\approx 0.1$ at $\sqrt{s} = 7$ TeV and $\approx 0.2$ at $\sqrt{s} = 14$ TeV, indicating
 a small but non-negligible contribution from the DPS channel to the total amount of $b\bar{b}$ pairs detected in LHCb. For comparison, the ratio `` cc/c '' assumes
 the value $\approx 0.06$ at $\sqrt{s} = 7$ TeV and $\approx 0.1$ at $\sqrt{s} = 14$ TeV.

\begin{figure}[t]
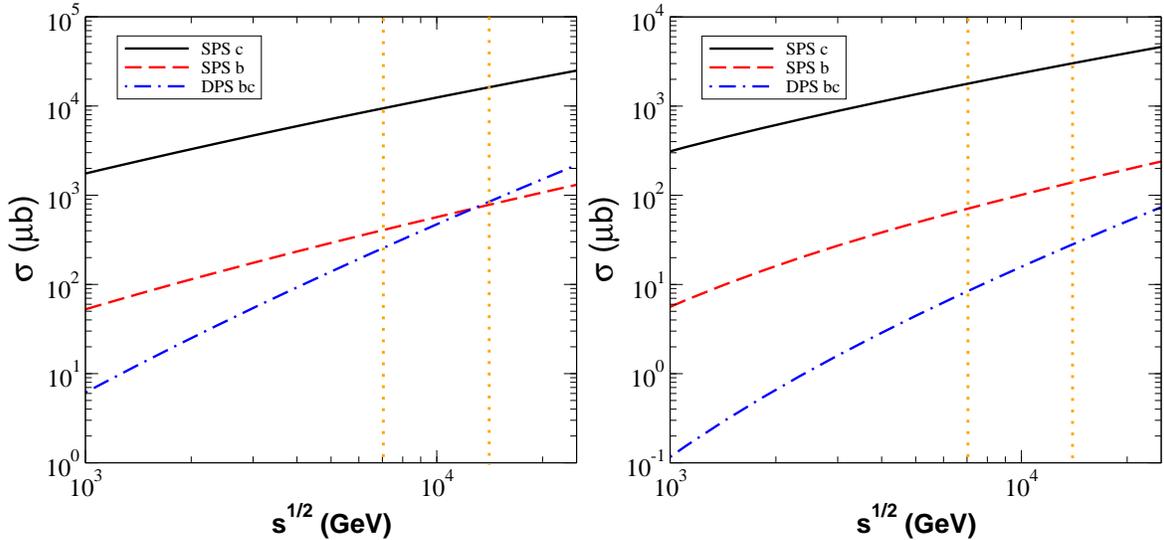

\begin{tabular}{cc}
\includegraphics[scale=0.45]{sps_lhc.eps}
\includegraphics[scale=0.45]{sps_lhcb.eps}
\end{tabular}
\caption{(Color online) $c\bar{c}$ and $b\bar{b}$ production cross sections in SPS events and $c\bar{c}b\bar{b}$ production cross section in DPS events as a 
function of the c.m.s. energy ($\sqrt{s}$) in the full LHC (left) and LHCb (right) rapidity range.}
\label{fig:5}
\end{figure}

\begin{figure}[t]
\begin{tabular}{cc}
\includegraphics[scale=0.40]{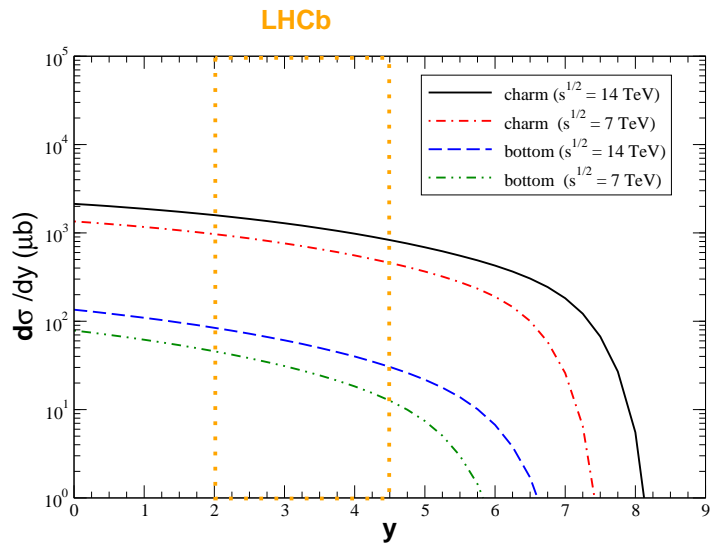}
\end{tabular}
\caption{(Color online) Differential cross sections as a function of the rapidity $y$ for SPS production of charm and bottom at the energies of LHC 
($\sqrt{s} =$ 7 TeV and $\sqrt{s} =$ 14 TeV).}
\label{fig:6}
\end{figure}

The behavior of the ratio `` bc/b '' can be better understood if we compare the energy dependence of the cross sections for the  SPS charm and bottom production 
with that predicted for the $c\bar{c}b\bar{b}$ production (denoted `` DPS bc'').
In Fig. \ref{fig:5} we present our predictions for these three different processes. As in the previous figure we present in the left panel  the results obtained 
integrating the cross sections in the full LHC rapidity range, while in the right panel the cross sections have been integrated in the LHCb rapidity range 
 $2 < y < 4.5$. In the first case, we can see that the  `` DPS bc '' prediction grows up more rapidly with the energy than those corresponding to the SPS 
processes. This implies that  the `` DPS bc '' prediction becomes of the same order of the `` SPS b '' one. In contrast, if we consider the LHCb rapidity range,
 the energy dependence of the three processes are not very distinct, which implies that the `` DPS bc '' prediction is always smaller than the `` SPS b '' one. 
This conclusion comes from the different rapidity distributions for charm and bottom production, which are presented in  Fig. \ref{fig:6}, where now the two 
vertical dotted lines delimit the rapidity interval probed by the LHCb. We observe that by considering the limited interval $2 < y < 4.5$ we are taking
 only a fraction of the total amount of charm and bottom produced. The product of these cross sections, integrated in the rapidity interval of LHCb, is much 
smaller than the product of the cross sections integrated in the whole phase space. Moreover the differential cross section for bottom production
 falls down suddenly when $y>6$ while the charm production cross section presents the same behaviour only for $y>7.5$. Therefore, for $y>6$ the bottom production
 is negligible when compared with its production in the region $y<6$. On the other hand, the charm production is still abundant in the interval $6<y<7.5$, being
 negligible only for $y>7.5$. This extra contribution of charm production in the interval $6<y<7.5$, in which the bottom production is very small, 
is a second factor that makes the total cross section for $c\bar{c}b\bar{b}$ production in DPS much greater than the one obtained in the limited region $2<y<4.5$.

\section{Summary}
\label{conc}

The contribution of multiple parton scatterings in the LHC energy range is expected to be non-negligible due to the large number of low-$x$ gluons present in 
the incident hadrons. The high partonic density should modify the QCD dynamics introducing non-linear effects (with  the possible formation of a Color 
Glass Condensate) and should enhance the probability of having two or more hard interactions. In this paper we consider the production of double heavy quark
 pairs taken into account the saturation effects. 
We  estimated the ratio between the double and single parton scattering cross sections for the full rapidity range of the LHC and for the rapidity range of the 
LHCb experiment. The previous prediction  that for the charm production the double parton scattering contribution  becomes comparable with the single parton
 scattering one at  LHC energies has been confirmed. Moreover, we demonstrated that this result remains valid when one considers saturation effects in the 
calculations and 
 that the production of $c\bar{c}b\bar{b}$ contributes significantly for the bottom production.  Finally, we obtained that for the LHCb kinematical range the 
ratio is strongly reduced.
We have estimated the DPS contribution considering the simple factorized model, which implies that our predictions should be taken with some caution, especially 
in the kinematical range when this contribution is large. However, we believe that our predictions can be considered as a reasonable first approximation and  our
 study can motivate the experimental analysis of this particular final state and the theoretical development of more detailed analysis.

\begin{acknowledgments}
This work was  partially financed by the Brazilian funding
agencies FAPESP, CNPq, CAPES and FAPERGS.
\end{acknowledgments}

\end{document}